\documentclass[twoside]{dis07}
\usepackage[latin1]{inputenc}
\usepackage[dvips]{graphicx,epsfig,color}
\usepackage{wrapfig,rotating}
\usepackage{amssymb,amsmath,array}
\usepackage{cite}

\begin{document}
\title{ 
{\vspace{-1.5cm}\normalsize \sl  DESY 07-086 \hfill {\tt arXiv:0706.3297v1 [hep-ph]}}\\
\vspace{1.5cm}
Higher Mellin Moments for Charged Current DIS 
}

\author{M.~Rogal and S.~Moch
%
%
\vspace{.3cm}\\
%
Deutsches Elektronensynchrotron DESY \\
Platanenallee 6, D--15738 Zeuthen - Germany \\
%
}

\maketitle

\begin{abstract}
We report on our recent results for deep-inelastic neutrino($\nu$)-proton($P$) scattering. 
We have computed the perturbative QCD corrections to three loops 
for the charged current structure functions $F_2$, $F_L$ and $F_3$ 
for the combination $\nu P - \bar \nu P$.
In leading twist approximation we have calculated the first six odd-integer Mellin moments 
in the case of $F_2$ and $F_L$ and the first six even-integer moments in the case of $F_3$.
As a new result we have obtained the coefficient functions to ${\cal O}(\alpha_s^3)$ 
and we have found the corresponding anomalous dimensions to agree with known results in the literature.
\end{abstract}

\section{Introduction}
\label{sec:introductionMR}
In our recent research~\cite{url}, we extended the program of calculating higher order perturbative QCD corrections 
to the structure functions of charged current deep-inelastic scattering (DIS).
Our studies are motivated by the increasingly accurate measurements of 
neutral and charged current cross sections at HERA with a polarized beam of 
electrons and positrons~\cite{Chekanov:2003vw,Adloff:2003uh,Aktas:2005ju}.
At the same time we are also able to quantitatively improve predictions for physics 
at the front-end of a neutrino-factory, see e.g. Ref.~\cite{Mangano:2001mj}.
To be specific, we consider neutrino($\nu$)-proton($P$) scattering in the combination $\nu P - \bar \nu P$,
which corresponds to charged lepton-proton DIS as far as QCD corrections are concerned.
Following Refs.~\cite{Larin:1993vu,Larin:1996wd,Retey:2000nq,Moch:2001im,Blumlein:2004xt}
we compute the perturbative QCD (pQCD) predictions to three-loop accuracy for a number 
of fixed Mellin moments of the structure functions $F_2$, $F_L$ and $F_3$.

Within the framework of the operator product expansion (OPE), and working in Mellin space, 
$F_2^{ \nu P - \bar \nu P}$ and $F_L^{ \nu P - \bar \nu P}$ are functions of odd Mellin moments only, 
while only even moments contribute to $F_3^{\nu P -\bar \nu P}$. 
This is opposite to the case of the neutral current structure functions where only even Mellin moments 
contribute, and to the charged current case for $\nu P + \bar \nu P$ scattering~\cite{Moch:2007gx}, 
which is defined through the OPE for odd Mellin moments only.
In the latter results for $F_2^{ \nu P + \bar \nu P}$ and $F_L^{ \nu P + \bar \nu P}$ to three-loops 
can also be directly checked in electromagnetic DIS~\cite{Moch:2004xu,Vermaseren:2005qc} 
while parameterizations for $F_3^{\nu P + \bar \nu P}$ to three-loop accuracy are given in Ref.~\cite{Vogt:2006bt}.

\section{General  formalism}
\label{sec:formalism}
We consider unpolarized inclusive deep-inelastic lepton-nucleon scattering,
\begin{eqnarray}
\label{eq:dis}
  l(k) \:+\: {\rm nucl}(p) \:\:\rightarrow\:\: l^{\, \prime}(k^{\,\prime}) \:+\:  X\, ,
\end{eqnarray}
where $l(k),\, l^{\,\prime}(k^{\,\prime})$ are leptons of momenta $k$ and 
$k^{\, \prime}$, ${\rm nucl}(p)$ denotes a nucleon of momentum $p$ and 
$X$ stands for all hadronic states allowed by quantum number conservation.
In our research we are concentrating on charged current
neutrino($\nu$)-proton($P$) scattering, i.e. $\nu P$, $\bar \nu P$ via $W^{\pm}$ boson exchange.
As is well known, the differential cross section for the reaction~(\ref{eq:dis}) 
can be written as a product of leptonic $L_{\mu\nu}$ and hadronic $W_{\mu\nu}$ tensors
\begin{eqnarray}
\label{eq:diffcrosssec}
d \sigma \propto L^{\mu\nu} W_{\mu\nu}\, ,
\end{eqnarray}
with $L^{\mu\nu}$ for electroweak or pure electromagnetic gauge boson exchange
given in the literature, see e.g. Ref.~\cite{Yao:2006px}.
The hadronic tensor $W^{\mu\nu}$ in Eq.~(\ref{eq:diffcrosssec}) can be written 
in terms of so called structure functions $F_{i}$, $i=2,3,L$.

We are interested in the Mellin moments of structure functions, defined as 
\begin{eqnarray}
\label{eq:mellindefF2L}
\displaystyle
F_{i}(n,Q^2) &=&
\int\limits_0^1 dx\, x^{n-2} F_{i}(x,Q^2)\, ,\quad
i = 2,L\, 
\end{eqnarray}
and for $F_{3}(n,Q^2)$ one has similar relation with $n$ replaced by $n+1$ on the r.h.s. of Eq.~(\ref{eq:mellindefF2L}). 
Here $Q^2=-q^2 > 0,$ $q=k-k^{\,\prime}$ and $x$ is the 
Bjorken scaling variable defined as $x=Q^2/ (2p\cdot q)$ with $0 < x \leq 1$.

With the help of the optical theorem and Cauchy`s theorem from complex analysis one can 
relate the Mellin moments of structure functions to the parameters of the OPE 
for the nucleon forward Compton amplitude $T_{\mu\nu}$:
\begin{eqnarray}
  \label{eq:F2mellinMR}
F_{i}(n,Q^2)
  &=& 
  C_{i,{\rm ns}}\left(n,\frac{Q^2}{\mu^2},\alpha_s\right) 
  A_{ \rm nucl }^{\rm ns}\left(n,{\mu^2}\right)\, , 
  \quad\quad\quad i=2,3,L\,  
\end{eqnarray}
and the OPE for  $T_{\mu\nu}$ reads as 
\begin{eqnarray}
\label{eq:OPEnucl}
T_{\mu\nu} 
&=&
2 \sum_{n} {\omega}^n \biggl[ e_{\mu\nu}\, C_{L,{\rm ns}}\biggl(n,\frac{Q^2}{\mu^2},\alpha_s\biggr) 
     + d_{\mu\nu}\, C_{2,{\rm ns}}\biggl(n,\frac{Q^2}{\mu^2},\alpha_s\biggr)  
\nonumber\\
& &\mbox{}
+ {\rm{i}} \epsilon_{\mu\nu\alpha\beta} \frac{p^\alpha q^\beta}{p\!\cdot\! q} 
C_{3,{\rm ns}}\biggl(n,\frac{Q^2}{\mu^2},\alpha_s\biggr) \biggr] A_{{\rm{nucl}}}^{\rm ns}\left(n,{\mu^2}\right) 
+ {\rm{higher\,\, twists}}
\, ,
\end{eqnarray}
where higher twist contributions are omitted. $C_{i,{\rm ns}}$ denote the Wilson
coefficients which are calculable in pQCD and $A_{{\rm{nucl}}}^{\rm ns}$ are 
matrix elements of quark non-singlet operators. 
The latter are not calculable in pQCD, rather they have to be extracted from experimental data. 
We restrict ourselves to quark non-singlet (ns) operators only since only these
give nonvanishing contributions in the combination ${\nu P-\nu N}$ 
(see Ref.~\cite{Moch:2007gx} for details).
  
Eq.~(\ref{eq:F2mellinMR}) provides the basis to obtain Mellin moments of DIS structure functions 
in our approach by means of the OPE and the optical theorem.
Furthermore, from the careful examination of the symmetry properties of the 
forward Compton amplitude $T_{\mu\nu}$ and, related, the underlying Feynman diagrams, 
one can convince oneself that for the charged current ${\nu P-\nu N}$ DIS, 
one encounters functions of only odd $n$ for $F_2$ and $F_L$ and, 
functions of only even $n$ for $F_3$, respectively~\cite{Moch:2007gx}.

The pQCD calculation of Wilson coefficients $C_{i,{\rm ns}}$ proceeds through the following steps. 
From the first principles we calculate the partonic forward Compton amplitude $t_{\mu\nu}$.
The partonic equivalent of the OPE Eq.~(\ref{eq:OPEnucl}) for $t_{\mu\nu}$
contains the {\it same}  coefficients $C_{i,{\rm ns}}$ as in Eq.~(\ref{eq:OPEnucl}) 
and quark matrix elements $A_{{\rm{q}}}^{\rm ns}$. 
Projection on the $n$'th Mellin moment of 
OPE and on the $i$'th parton invariant ($i=2,3,L$) 
with the help of the operator ${\cal P}_{n,i}^{\mu\nu}$ we get
\begin{eqnarray}
\label{eq:TmunuPartonMomNS}
t_{i,{\rm{ns}}} \left(n,\frac{Q^2}{\mu^2},\alpha_s,\epsilon\right)
\equiv
{\cal P}_{n,i}^{\mu\nu}\, t_{\mu\nu}
=
C_{i,{\rm ns}}\left(n,\frac{Q^2}{\mu^2},\alpha_s,\epsilon\right) 
Z_{\rm{ns}}\left(\alpha_s,\frac{1}{\epsilon}\right)
A^{{\rm ns},{\rm tree}}_{\rm q}(n,\epsilon)  
\, .
\end{eqnarray}
Both sides of Eq.~(\ref{eq:TmunuPartonMomNS}) are renormalized.
In particular the renormalization of the local quark operator matrix element
$A_{{\rm{q}}}^{\rm ns}$ gives rise to the factor $Z_{\rm{ns}}$ on the r.h.s. of Eq.~(\ref{eq:TmunuPartonMomNS}).
This equation is our starting point for an iterative determination of the coefficient functions 
$C_{i,{\rm ns}}$ and the anomalous dimension $\gamma_{\rm ns}$. 
The latter appears in a series expansion of $Z_{\rm ns}$ in powers of the strong coupling $\alpha_s$ 
and negative powers of the parameter $\epsilon$ of dimensional regularization, $D=4-2\epsilon$.
The $C_{i,{\rm ns}}$ on the other hand are expanded in $\alpha_s$ and in positive powers of $\epsilon$.
Thus the l.h.s. of Eq.~(\ref{eq:TmunuPartonMomNS}) leads to a well defined
determination of $C_{i,{\rm ns}}$ and $Z_{\rm ns}$ in pQCD.

\section{Calculation and checks}
\label{sec:calculation}
In the previous section, we have briefly explained the method to obtain 
Mellin moments of the DIS charged current structure functions 
$F_2^{\nu P -\bar \nu P}$, $F_3^{\nu P -\bar \nu P}$ and $F_L^{\nu P -\bar \nu P}$ 
together with their respective coefficient functions and anomalous dimensions.
To that end we have calculated the Lorentz invariants of the parton Compton
amplitude $t_{i,{\rm ns}}$ , $i=2,3,L$, as given in the l.h.s. of Eq.~(\ref{eq:TmunuPartonMomNS}). 
Due to the large number of diagrams involved in the calculations up to order $\alpha_{s}^{3}$ 
sufficient automatization is necessary.
First of all, we have generated 3633 diagrams up to three loops with the program {\sc Qgraf}~\cite{Nogueira:1991ex}.
For all further calculations we have relied on the latest version of 
the symbolic manipulation program {\sc Form}~\cite{Vermaseren:2002rp,Vermaseren:2006ag}.

For the treatment of {\sc Qgraf} output, such as analysis of the 
topologies, the explicit implementation of Feynman rules etc. 
we have adapted a dedicated {\sc Form} procedure {\it conv.prc} from 
previous work, e.g. Ref.~\cite{Vermaseren:2005qc}. 
Most importantly, this procedure tries to exploit as many symmetry properties 
of the original Feynman diagrams as possible in order to reduce their total number.
  
For the calculation of the color factors for each Feynman diagram 
we have used the {\sc Form} package {\it color.h}~\cite{vanRitbergen:1998pn}.
The actual calculation of the Mellin moments of the Feynman integrals
has made use of the {\sc Form} version of {\sc Mincer}~\cite{Larin:1991fz}. 
Finally, on top of {\sc Mincer} and {\sc Minos}~\cite{Larin:1996wd} some shell scripts
managed the automatic runs of both programs for different parts of the calculation. 

We have performed various checks on our computation. 
Most prominently, we have kept all powers of the gauge parameter $\xi$ throughout 
the entire calculation for Mellin moments $n\leq 10$ to check that any $\xi$-dependence vanishes in our final results.
The Mellin moments with $n>10$ were calculated without gauge parameter to
facilitate the computations which become increasingly more  complicated for higher Mellin $n$ values. 
For these moments we also used {\sc TForm}~\cite{Tentyukov:2007mu}, the multi-threaded version of {\sc Form}.
On machines with multi-core processors this leads to a significant speed up of
our calculations, e.g. a speed-up of $\simeq 5$ on a two-core four processor machine.

We agree with the literature as far as the 
two-loop coefficient functions~\cite{vanNeerven:1991nn,Zijlstra:1991qc,Zijlstra:1992kj,Zijlstra:1992qd,Moch:1999eb} 
and the three-loop anomalous dimensions~\cite{Moch:2004pa} are concerned.
In addition, for the first Mellin moment of the coefficient function $C_{2,\rm ns}$ 
we have obtained exactly $C_{2,\rm ns} = 1$  
to all orders in $\alpha_s$ which is in agreement with the Adler sum rule for DIS structure functions,
\begin{equation}
  \label{eq:adler-sumrule}
  \int\limits_0^1 \frac{dx}{x} \biggl(F_2^{\nu P}(x,Q^{2}) - F_2^{\nu N}(x,Q^{2}) \biggr) = 2 
\, .
\end{equation}
The Adler sum rule measures the isospin of the nucleon in the quark-parton model and does
not receive any perturbative or non-perturbative corrections in QCD, 
see e.g. Ref.~\cite{Dokshitzer:1995qm}.
Therefore, this result is another important check of the correctness of our results.

\section{Conclusions}
\label{sec:conclusions}
We have reported on new results for Mellin moments 
of the charged current DIS structure functions $F_2^{\nu P - \bar \nu P}$, 
$F_L^{\nu P - \bar \nu P}$ and $F_3^{\nu P - \bar \nu P}$ including 
the perturbative QCD corrections to three loops.
In the former case ($F_2$, $F_L$) we have computed the first six odd-integer Mellin moments 
while in the latter case ($F_3$), the first six even-integer moments have been obtained.
The results for $F_{2,L}^{\nu P - \bar \nu P}$ $n=1,3,5,7,9$ and for $F_3^{\nu P - \bar \nu P}$ $n=2,4,6,8,10$ 
are available in Ref.~\cite{Moch:2007gx}. 
Results for $n=11$ in the former case and for $n=12$ in the latter will be published elsewhere.
Finally, the discussion of phenomenological consequences of our Mellin space
results along with approximate parameterizations the coefficient functions in $x$ 
are deferred to Ref.~\cite{MRV1}.

\begin{footnotesize}

\end{footnotesize}



\begin{thebibliography}{10}

\bibitem{url}
Slides:\\
  \verb$http://indico.cern.ch/contributionDisplay.py?contribId=24&sessionId=14&confId=9499$.

\bibitem{Chekanov:2003vw}
S.~Chekanov et~al.
\newblock {\em Eur. Phys. J.}, C32:1--16, 2003.

\bibitem{Adloff:2003uh}
C.~Adloff et~al.
\newblock {\em Eur. Phys. J.}, C30:1--32, 2003.

\bibitem{Aktas:2005ju}
A.~Aktas et~al.
\newblock {\em Phys. Lett.}, B634:173--179, 2006.

\bibitem{Mangano:2001mj}
M.~L. Mangano et~al.
\newblock hep-ph/0105155, 2001.

\bibitem{Larin:1993vu}
S.~A. Larin, T.~van Ritbergen, and J.~A.~M. Vermaseren.
\newblock {\em Nucl. Phys.}, B427:41--52, 1994.

\bibitem{Larin:1996wd}
S.~A. Larin, P.~Nogueira, T.~van Ritbergen, and J.~A.~M. Vermaseren.
\newblock {\em Nucl. Phys.}, B492:338--378, 1997.

\bibitem{Retey:2000nq}
A.~Retey and J.~A.~M. Vermaseren.
\newblock {\em Nucl. Phys.}, B604:281--311, 2001.

\bibitem{Moch:2001im}
S.~Moch, J.~A.~M. Vermaseren, and A.~Vogt.
\newblock {\em Nucl. Phys.}, B621:413--458, 2002.

\bibitem{Blumlein:2004xt}
J.~Bl\"umlein and J.~A.~M. Vermaseren.
\newblock {\em Phys. Lett.}, B606:130--138, 2005.

\bibitem{Moch:2007gx}
S.~Moch and M.~Rogal.
\newblock arXiv:0704.1740 [hep-ph], 2007.

\bibitem{Moch:2004xu}
S.~Moch, J.~A.~M. Vermaseren, and A.~Vogt.
\newblock {\em Phys. Lett.}, B606:123--129, 2005.

\bibitem{Vermaseren:2005qc}
J.~A.~M. Vermaseren, A.~Vogt, and S.~Moch.
\newblock {\em Nucl. Phys.}, B724:3--182, 2005.

\bibitem{Vogt:2006bt}
A.~Vogt, S.~Moch, and J.~Vermaseren.
\newblock {\em Nucl. Phys. Proc. Suppl.}, 160:44--50, 2006.

\bibitem{Yao:2006px}
W.~M. Yao et~al.
\newblock {\em J. Phys.}, G33:1--1232, 2006.

\bibitem{Nogueira:1991ex}
P.~Nogueira.
\newblock {\em J. Comput. Phys.}, 105:279--289, 1993.

\bibitem{Vermaseren:2002rp}
J.~A.~M. Vermaseren.
\newblock {\em Nucl. Phys. Proc. Suppl.}, 116:343--347, 2003.

\bibitem{Vermaseren:2006ag}
J.~A.~M. Vermaseren and M.~Tentyukov.
\newblock {\em Nucl. Phys. Proc. Suppl.}, 160:38--43, 2006.

\bibitem{vanRitbergen:1998pn}
T.~van Ritbergen, A.~N. Schellekens, and J.~A.~M. Vermaseren.
\newblock {\em Int. J. Mod. Phys.}, A14:41--96, 1999.

\bibitem{Larin:1991fz}
S.~A. Larin, F.~V. Tkachov, and J.~A.~M. Vermaseren.
\newblock NIKHEF-H-91-18.

\bibitem{Tentyukov:2007mu}
M.~Tentyukov and J.~A.~M. Vermaseren.
\newblock hep-ph/0702279, 2007.


\bibitem{vanNeerven:1991nn}
W.~L. van Neerven and E.~B. Zijlstra.
\newblock {\em Phys. Lett.}, B272:127--133, 1991.

\bibitem{Zijlstra:1991qc}
E.~B. Zijlstra and W.~L. van Neerven.
\newblock {\em Phys. Lett.}, B273:476--482, 1991.

\bibitem{Zijlstra:1992kj}
E.~B. Zijlstra and W.~L. van Neerven.
\newblock {\em Phys. Lett.}, B297:377--384, 1992.

\bibitem{Zijlstra:1992qd}
E.~B. Zijlstra and W.~L. van Neerven.
\newblock {\em Nucl. Phys.}, B383:525--574, 1992.

\bibitem{Moch:1999eb}
S.~Moch and J.~A.~M. Vermaseren.
\newblock {\em Nucl. Phys.}, B573:853--907, 2000.

\bibitem{Moch:2004pa}
S.~Moch, J.~A.~M. Vermaseren, and A.~Vogt.
\newblock {\em Nucl. Phys.}, B688:101--134, 2004.

\bibitem{Dokshitzer:1995qm}
Y.~L. Dokshitzer, G.~Marchesini, and B.~R. Webber.
\newblock {\em Nucl. Phys.}, B469:93--142, 1996.


\bibitem{MRV1}
S.~Moch, M.~Rogal, and A.~Vogt.
\newblock DESY 07-048 , 2007.

\end{thebibliography}
\end{document}